# Pneumatic Pressure Cell with Twin Diaphragms Embedding Spherical Corrugations in a Dual Diaphragm Structure

A. Cellatoglu[1] and K. Balasubramanian[2]

[1]Department of Computer Engineering, European University of Lefke
Turkish Republic of Northern Cyprus, Mersin 10, TURKEY

[2] Department of EE Engineering, European University of Lefke
Turkish Republic of Northern Cyprus, Mersin 10, TURKEY

### Abstract

Thin metallic shallow spherical diaphragms are being used for measuring pneumatic pressure in process industries. The drift in vertex realized due to application of pressure is transformed into electrical signal and this is calibrated for pressure. We now propose a modified structure for the pressure cell by having double ended shallow spherical shells embedded with spherical corrugations as to enhance the sensitivity to a greater extent. By having dual such installation in the structure of the pressure cell it concedes further increase in sensitivity. The construction details of the diaphragm structure, theory and analysis to assess the performance are presented.

Keywords: *Dual Diaphragm Structure, Spherical Corrugations, Pressure Cell, Sensitivity Enhancement, Twin Diaphragms*

## 1. Introduction

Pneumatic pressure cells employing diaphragms are conventionally being used in process industries. Most diaphragm based cells use strain gauge pickups[1] and certain cells use vibrating wires[2]. Thin metallic diaphragms yield relatively larger drift in the vertex when pressure is acting on it. This drift in diaphragm was transformed into electrical signal by using inductive pickup, capacitive pickup and LVDT pickups[3-5]. Enhancing further the sensitivity a structure with dual diaphragm structures were also reported[6,7]. A geometrical structure involving corrugations along with a diaphragm for making a pressure cell was also reported[8]. Yielding further enhancement in sensitivity we propose now dual diaphragm structure for the pressure cell with each one having twin diaphragms with embedded corrugations. This enhances the sensitivity to a greater extent compared to other types diaphragm cells. Evidently, the sensitivity enhancement permits the cell to measure weaker pressure signals and makes it more useful.

### 1.1 Thin Metallic Shallow Spherical Diaphragm

The drift in the vertex of a thin diaphragm with shallow spherical structure depends on the elastic properties of the materials used for the diaphragm, its geometry and size. With an appropriate pickup the enhancement of the drift produces significant raise in the level of signal. The relationship between the applied pressure and the drift in the vertex is established by solving governing equations with boundary conditions of the diaphragm. Fig.1 shows the schematic of a shallow spherical shell inflated by pneumatic pressure wherein $w$ denotes the altitude of a point in the surface of shell measured from plane of the rim at a radial distance $r$. The altitude of the vertex from the rim plane is denoted as $f$. The altitude $f$ is shown to be dependent on pressure and various mechanical parameters [9-11] as

$$f = A - \frac{ra}{3.A} \qquad (1)$$

where *ra:* radius of the rim of the diaphragm

$$A = \left(\frac{\eta}{2} + \varsigma\right)^{1/3} \qquad (2)$$

$$\varsigma = \left(\frac{\alpha^3}{27} + \frac{\eta^2}{4}\right)^{1/2} \qquad (3)$$

$$\alpha = \frac{56.h^2}{(1+\gamma)(23-9.\gamma)} \qquad (4)$$

$h$ : plate thickness and $\gamma$ : Poisson ratio.

$$\eta = \frac{7.p.ra^4 h^2}{8.D(1+\gamma)(23-9.\gamma)} \qquad (5)$$

where $D$ : flexural rigidity which is a mesurement of stiffness and is given by

$$D = \frac{E.h^3}{12.(1-\gamma^2)} \quad (6)$$

where $E$ : Young's modulus.

The altitude of the diaphragm $w$ was shown to be related to $f$ [7],[9] as given below

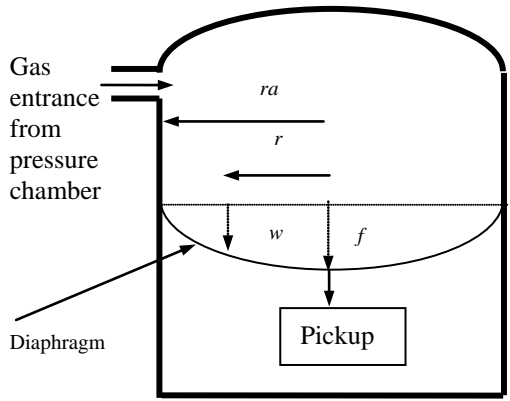

Fig. 1 Schematic of Diaphragm based Pressure Cell

$$w(r) = f.\left(1-\left(\frac{r}{a}\right)^2\right)^2 \quad (7)$$

The pickup installed in the diaphragm should be efficient in extracting the drift information into electrical signal. One of its essential requirement is that it should be light in weight such that it will not impose any payload which would affect the mechanical properties of the diaphragm. Therefore, in the past weightless inductive pickups and capacitive pickups were attempted. Now for the proposed twin dual diaphragm structure we use modified hollow cylindrical capacitive pickup as it adapts better to this geometry.

## 2. Twin Dual Diaphragms With Embedded Corrugations

### 2.1 Geometry and Structure

A simplified schematic of the geometrical structure illustrating the principle of the corrugations embedded twin dual diaphragms installed with capacitive pickup is shown in Fig.2. This has improved geometrical structure compared to the earlier pressure cell[8] wherein four spherical diaphragms with spherically embedded corrugations are employed. Each diaphragm structure of the twin diaphragms has two diaphragms with multiple corrugations embedded. The diaphragms are spaced facing symmetrically opposite to each other wherein the pneumatic pressure is allowed in at the vertex position of the first diaphragm and capacitor plates for the pickup are extended in the other diaphragm.

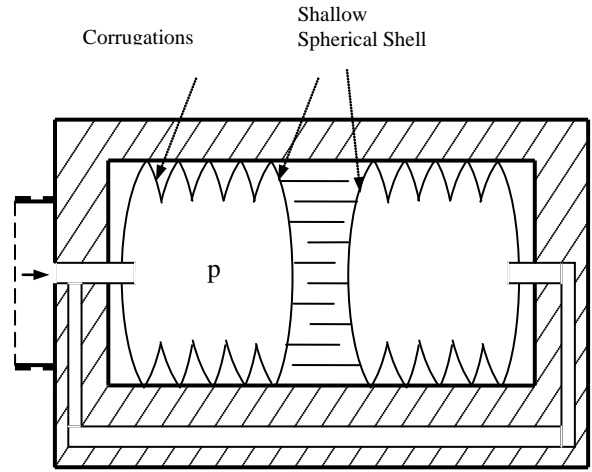

Fig.2. Schematic of he Sectional View of the Double ended Diaphragm Structure with Embedded Corrugations

An array of light weight Silver coated Teflon cylindrical plates arranged in one diaphragm slide through the gaps between the similar cylindrical plates arranged in the other diaphragm. When pressure is applied it causes both corrugated diaphragms to move in opposite directions. When the plates move in the gaps of each other it results in change of effective area of plates contributing to change in capacitance. By classical theory, the capacitance between any two plates is given by

$$C = \frac{\varepsilon A}{d} \quad (8)$$

where  $\varepsilon$ : Dielectric constant,
  A: Area of plates and
  d: displacement between plates.

The instantaneous capacitance is detected and converted into voltage for measurement and transmission to any other location for processing and display.

### 2.2 Drift Contributions from Spherical Corrugations

The purpose of embedding corrugations is to yield relatively larger drift in vertex due to application of pressure. For assessment of the drift in vertex contributed by a shallow spherical corrugation, the geometry of a segment of corrugation shown in Fig.3 is referred.

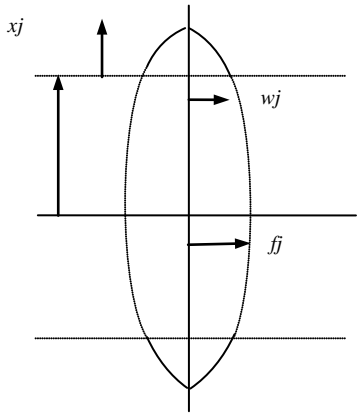

Fig.3. A Segment of Shallow Spherical Corrugation

The corrugation with groove length $xj$ exhibits the axial distance $wj$ at the point of terminating the groove where as the axial distance of vertex passing through the centre of the diaphragm is $fj$.

Knowing $fj$, the distance $wj$ can be obtained by using the relation (7) with $r$ made as $rj$ and $a$ made as $ra$.

Therefore, each slanted sector of a groove contributes an axial distance of $wj$ for producing drift when pressure acts on the diaphragm structure.

For standard applied pressure $ps$, the end spherical shell of the diaphragm yields the drift in the vertex $fs$ governed by (1) to (6)

For an unknown pressure $px$, if the drift in vertex is denoted as $fx$, then change in drift of the vertex $\Delta f$ is given by

$$\Delta f = fx - fs \qquad (9)$$

This change in drift of the vertex $\Delta f$ is experienced due to change in pressure $\Delta p = px - ps$.

The axial distance experienced at any instant due to application of instantaneous pressure $px$ acting on the diaphragms is obtained from the contributions of all slanted sectors of the corrugations and from the drifts of the ending up shallow spherical shells.

If one sector of corrugation contributes an axial distance of $wx$ due the applied pressure $px$, then the axial distance materialized by one diaphragm with $n$ sectors of corrugation is

$$fn = n.wx \qquad (10)$$

There are two diaphragms at the end. Including the displacement contributed by two ending diaphragms the resulting drift is

$$fd = fn + 2.fx \qquad (11)$$

Since there are two corrugated diaphragms the relative displacement between the two vertices $fdr$ is given by

$$fdr = 2.fd \qquad (12)$$

This distance $fdr$ realized is much greater than that of the distance realized from a single diaphragm $f$. Therefore, the application of pressure $px$ causes relatively larger capacitance with the array of capacitor plates. As a result, the sensitivity contributed by dual corrugated twin diaphragm based pressure cell is exceedingly larger than that of single diaphragm pressure cell. Fig.4 shows the relationship between the capacitances realized in dual corrugated diaphragm cell compared to that of dual diaphragm cell and single diaphragm cell.

The diaphragm used is a medium strength Al alloy of 2.5cm radius, 0.6mm thickness and 0.5cm of height for the spherical shell with Young's modulus 200GN/m$^2$ and Poisson ratio 0.3. Corrugations with 8 sectors are employed in each diaphragm structure. Silver coated thin Teflon cylindrical plates with radius of outer cylinder as 2cm and length 1.5cm is employed as capacitor plates. The length of the innermost plate used is 1.2 cm. With the application of pressure of 10 Pascals the drift in the vertices of diaphragms has shown a capacitance of 15pF where as when the pressure is void the capacitance developed is 2PF. For a similar arrangement of capacitor plates in dual diaphragm pressure cell excluding corrugations the capacitance resulted is 7pF for the pressure of 10 Pascals. For a single diaphragm cell the capacitance realized is nearly half compared to dual diaphragm cell.

With an application of pressure of 10 Pascals to the cell with 8 sectors of corrugations it has resulted a capacitance of 15 pF and if number of sectors decrease then the capacitance also would decrease accordingly as per the relationship already discussed.

For the sector geometry of $xj$ as 0.5cm, $rj$ as 2 cm the $fj$ resulted is 0.6 cm and $wj$ is 0.2078cm. The drift contribution per sector is 0.0776 cm for the applied pressure of 10 Pascals. Therefore, the total drift resulted by 8 sectors would be 0.622 cm. The contribution of drift due to two ending diaphragms would be 0.4cm. This has a total displacement of 1.022cm contributed by corrugations and diaphragms. Obviously the drift contribution reduces with the number of sectors employed and for four sectors this would be 0.711cm. Fig.5 shows the realization of capacitance as a function of number of sectors present in the corrugations. The pressure applied is kept as a reference of 10 Pascal's in all computations.

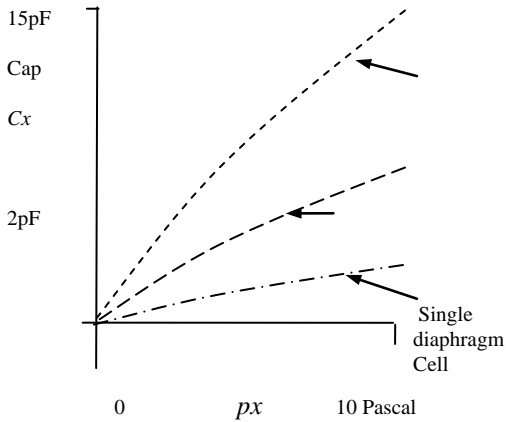

Fig.4. Capacitances Resulted in Different Diaphragm Structures

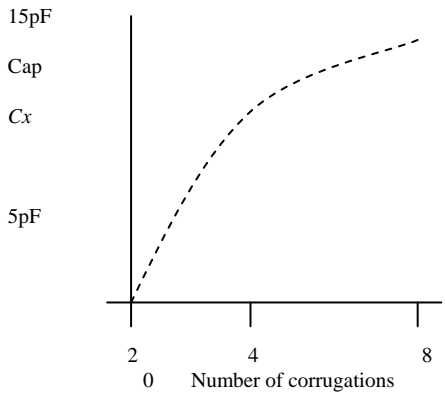

Fig.5. Capacitance *Cx* vs Corrugations

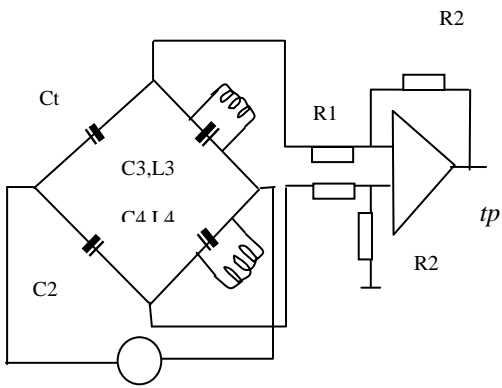

Fig.6. Transducer Circuit Schematic for Deriving Pressure Signal

### 2.3 Generation of Electrical Signal with Applied Pressure

The capacitance realized in the chamber of twin dual diaphragm pressure cell has to be transformed into electrical signal for transmitting, processing and display. Fig.6 shows the circuit schematic of the transducer producing the electrical signal. An AC bridge comprising the transducer capacitance *Ct*, lumped capacitor *C2*, parallel circuits of *C3,L3* and *C4,L4* is excited by a sinusoidal signal of frequency 10KHz. The error signal is picked up in the other two nodes of the bridge and is amplified by a difference amplifier. The amplified signal represents the intensity of the pressure signal. When the bridge is balanced the output signal would be zero. Whenever *Ct* changes due to change in pressure, then the unbalance in bridge gives error signal denoting the pressure applied.

The condition for balancing the bridge is as follows.

$$\frac{Ct}{C2} = \frac{(1-\varpi^2 L4.C4)}{(1-\varpi^2 L3.C3)} \quad (13)$$

The voltage gain *G* of the difference amplifier is given by

$$G = \left(R2/R1\right) \quad (14)$$

and the output voltage of the amplifier is

$$tp = G.ei \quad (15)$$

where *ei* is the error signal obtained from the bridge. These circuit parameters are chosen to amplify the signal to the required level against the variations due to environmental conditions such as temperature.

The elements of the bridge and the gain of the difference amplifier are selected such that desirable range of error signal is reached for optimized measurement with the type of diaphragms involved. There is an inherent nonlinearity involved in the transducer and this linearized and calibrated to pressure by using a circuit scheme shown in Fig.7. The analogue output from the differential amplifier of the transducer circuit is in AC form with a frequency of 10KHz and this is converted into DC by a standard precision rectifier using op-amps. This is then converted into binary form of data by using a 12-bit Analogue to Digital Converter (ADC) and driven with the address bus of the 4K EPROM where the lookup table of the linearized pressure data (*Pd*) is already stored. This binary data obtained at the addressed location is converted into analogue form by a Digital to Analogue Converter (DAC) and the calibrated pressure signal (*pa*) is sent externally for meeting other purposes. Also, the binary information from 4K EPROM is converted into BCD form again using lookup table approach implemented with EPROMs and driven to a display system for the display of pressure information. The display information is latched and updated every second automatically.

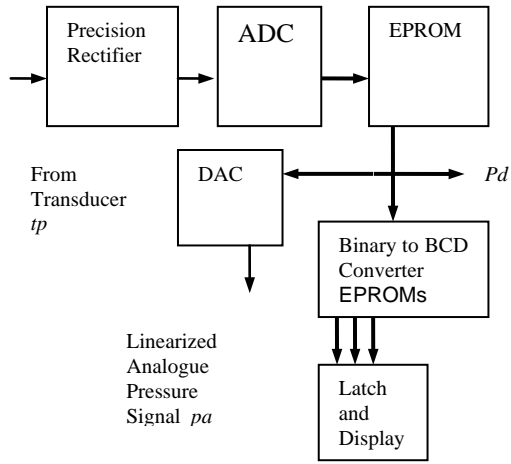

Fig.7 Schematic of Linearizer
Pd: Linearized Digital Pressure Data

## 3. Experimental Analysis

In order to assess the performance of the pressure cell both simulated and practical experiments are conducted.

### 3.1 Simulated Experiment

A pressure cell of twin dual diaphragms embedded with corrugations of 8 sectors shown in Fig.2 and its geometry and materials explained in section 2.2 is experimented by computing the capacitance realized for different pressure inputs covering its operating range. First the relative drift in the vertices of two diaphragms is computed as per relations (1) to (7) and (9) to (12). After then capacitance is computed as per relation (8). Lookup table approach is used in the computation wherever possible. Incidentally when we change the geometry and materials for the diaphragms, then lookup tables have to be changed accordingly. After computing the capacitance the error signal in the transducer is computed and the peak value is taken as the rectified voltage. It is again given the lookup table in the EPROM as to obtain the measured pressure signal. In this computation it has been found that the computed pressure tallies closely with the assumed pressure input.

### 3.2 Setup for Practical Experimentation

A microprocessor based experimental setup is realized for conducting practical experiments with the twin dual diaphragm pressure cell. Its simplified schematic is shown in Fig.8. The microprocessor is supported by RAM and EPROM memory and standard peripherals such as keyboard and display. With this scheme any type of pressure waveform could be generated and applied to the pressure cell and response could be read and analyzed.

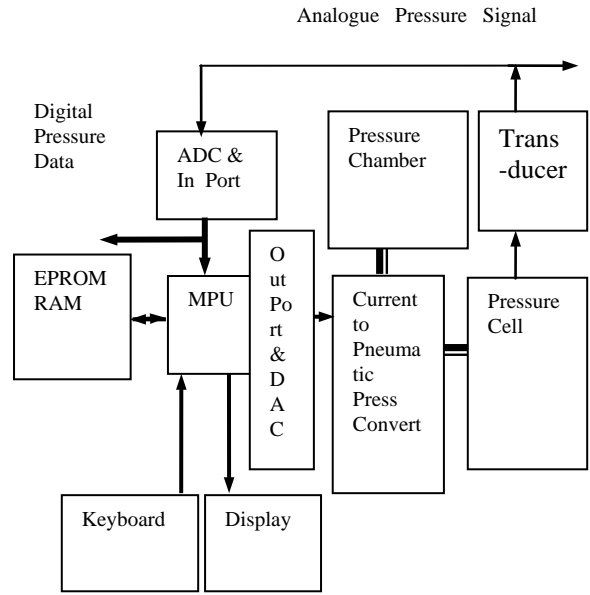

Fig.8. Experimental Setup

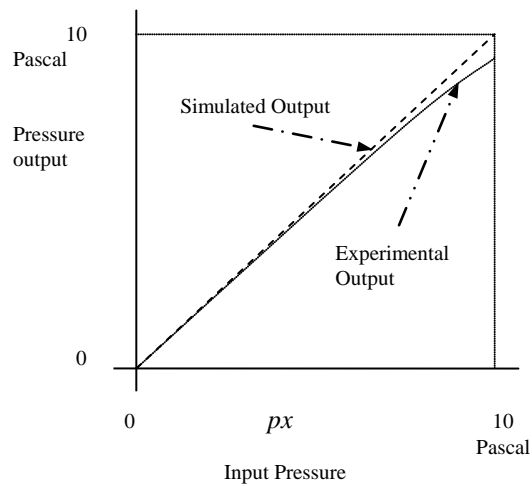

Fig.9. Experimental Results

The instantaneous analog voltage derived from a byte sent to the output port is converted into equivalent current and driven to a current-to-pneumatic pressure converter. This converter has a pressure chamber backup with compressed air and lets the output pressure at regulated level as set by the magnitude of the actuating current. Consequently the response in analogue form from the pressure cell is digitized and read by the microprocessor for further analysis. The data are saved in memory for further usage in making graphics plot and driving dot-matrix display.

While there is one to one correspondence between the applied pressure and simulated output pressure determined by computation there is a little difference in the practically obtained pressure especially at higher ranges of the applied pressure. Fig.9 shows both the results for different values of the applied pressure.

## 4. Conclusıon

The geometry of the capacitance pickup is optimized such as to produce maximum capacitance generated within the available space of the chamber. The plates of the capacitor are geometrically constructed in the form of concentric hollow cylinders. The cylindrical plates attached to one diaphragm would make movement within the gaps of cylindrical plates attached to another diaphragm and this would produce relatively more capacitance than the earlier parallel plate capacitors. When the sector length involved in corrugation is more it contributes more axial drift and the capacitance. Nevertheless, after some increase in sector length the geometrical stability of the dual diaphragm structure is affected and additional care should be taken to guide the path of the diaphragms when dynamic pressure is applied. Therefore for every diaphragm size there exists an optimum sector length and this is followed in designing the diaphragm structure. Twin diaphragm based pressure cell with spherical corrugations yield more axial displacement of the vertex contributing to generation of increased signal strength. Added to this effect, the dual diaphragm configuration enhances the capacitance output nearly two folds. The capacitance enhancement contributes to increased sensitivity in pressure measurement and promises to be more useful in several process industries.